# Derivation of a low multiplicative complexity algorithm for multiplying hyperbolic octonions

Aleksandr Cariow, Galina Cariowa, Jarosław Knapiński

Faculty of Computer Sciences and Information Technologies, Żołnierska 52
71-210 Szczecin, Poland

{atariov, gtariova, jknapinski}@wi.zut.edu.pl
tel. +48 91 4495573

**Abstract.** We present an efficient algorithm to multiply two hyperbolic (countercomplex) octonions. The direct multiplication of two hyperbolic octonions requires 64 real multiplications and 56 real additions. More effective solutions still do not exist. We show how to compute a product of the hyperbolic octonions with 26 real multiplications and 92 real additions. During synthesis of the discussed algorithm we use the fact that product of two hyperbolic octonions may be represented as a matrix–vector product. The matrix multiplicand that participates in the product calculating has unique structural properties that allow performing its advantageous factorization. Namely this factorization leads to significant reducing of the computational complexity of hyperbolic octonions multiplication.

## 1. Introduction

The development of theory and practice of data processing as well as necessity of solving more and more complex problems of theoretical and applied computer science requires using advanced mathematical methods and formalisms. At present hypercomplex numbers [1] are seeing increased application in various fields of digital signal and image processing [2-5], computer graphics and machine vision [6, 7], telecommunications [8-10] and in public key cryptography [11].

Among other arithmetical operations in the hypercomplex algebras, multiplication is the most time consuming one. The reason for this is, because the usual multiplication of these numbers requires $N(N-1)$ real additions and $N^2$ real multiplication. It is easy to see that the increasing of dimension of hypernumber increases the computational complexity of its multiplication. Therefore, reducing the computational complexity of the multiplication of hypercomplex numbers is an important theoretical and practical task. Efficient algorithms for the multiplication of various hypercomplex numbers already exist [12-21]. No such algorithms for the multiplication of the hyperbolic octonions have been proposed. In this paper, an efficient algorithm for this purpose is suggested.

## 2. Preliminary Remarks

A hyperbolic octonion can be defined as follows [22, 23]:

$$\hat{o} = b_0 + b_1 e_1 + b_2 e_2 + b_3 e_3 + b_4 \varepsilon_4 + b_5 \varepsilon_5 + b_6 \varepsilon_6 + b_7 \varepsilon_7 \qquad (1)$$

where $\{b_i\}, i=0,1,...,7$ are real numbers, $e_1$, $e_2$, $e_3$ are quaternion imaginary units, $\varepsilon_4$ ($\varepsilon_4^2 = 1$) is a counterimaginary unit, and the bases of hyperbolic octonions are defined as follows: $e_1\varepsilon_4 = \varepsilon_5$, $e_2\varepsilon_4 = \varepsilon_6$, $e_3\varepsilon_4 = \varepsilon_7$, ($\varepsilon_5^2 = \varepsilon_6^2 = \varepsilon_7^2 = 1$) [23]. The bases of hyperbolic octonions have multiplication rules as in Table 1:

Table 1. Rules for multiplication of hyperbolic octonion bases

| × | $e_1$ | $e_2$ | $e_3$ | $\varepsilon_4$ | $\varepsilon_5$ | $\varepsilon_6$ | $\varepsilon_7$ |
|---|---|---|---|---|---|---|---|
| $e_1$ | $-1$ | $e_3$ | $-e_2$ | $\varepsilon_5$ | $\varepsilon_4$ | $-\varepsilon_7$ | $\varepsilon_6$ |
| $e_2$ | $-e_3$ | $-1$ | $e_1$ | $\varepsilon_6$ | $\varepsilon_7$ | $\varepsilon_4$ | $-\varepsilon_5$ |
| $e_3$ | $e_2$ | $-e_1$ | $-1$ | $\varepsilon_7$ | $-\varepsilon_6$ | $\varepsilon_5$ | $\varepsilon_4$ |
| $e_4$ | $-\varepsilon_5$ | $-\varepsilon_6$ | $-\varepsilon_7$ | $1$ | $e_1$ | $e_2$ | $e_3$ |
| $e_5$ | $-\varepsilon_4$ | $-\varepsilon_7$ | $\varepsilon_6$ | $-e_1$ | $1$ | $e_3$ | $-e_2$ |
| $e_6$ | $\varepsilon_7$ | $-\varepsilon_4$ | $-\varepsilon_5$ | $-e_2$ | $-e_3$ | $1$ | $e_1$ |
| $e_7$ | $-\varepsilon_6$ | $\varepsilon_5$ | $-\varepsilon_4$ | $-e_3$ | $e_2$ | $-e_1$ | $1$ |

Assume we want to compute the product of two hyperbolic octonions $\hat{o}_3 = \hat{o}_1 \hat{o}_2$:

$$\hat{o}_1 = x_0 + x_1 e_1 + x_2 e_2 + x_3 e_3 + x_4 \varepsilon_4 + x_5 \varepsilon_5 + x_6 \varepsilon_6 + x_7 \varepsilon_7,$$

$$\hat{o}_2 = b_0 + b_1 e_1 + b_2 e_2 + b_3 e_3 + b_4 \varepsilon_4 + b_5 \varepsilon_5 + b_6 \varepsilon_6 + b_7 \varepsilon_7,$$

$$\hat{o}_3 = y_0 + y_1 e_1 + y_2 e_2 + y_3 e_3 + y_4 \varepsilon_4 + y_5 \varepsilon_5 + y_6 \varepsilon_6 + y_7 \varepsilon_7$$

Using "pen and paper" method we can write:

$$\begin{aligned}
\hat{o}_3 &= a_0 b_0 + a_0 b_1 e_1 + a_0 b_2 e_2 + a_0 b_3 e_3 + a_0 b_4 \varepsilon_4 + a_0 b_5 \varepsilon_5 + a_0 b_6 \varepsilon_6 + a_0 b_7 \varepsilon_7, \\
&+ a_1 b_0 e_1 + a_1 b_1 e_1^2 + a_1 b_2 e_1 e_2 + a_1 b_3 e_1 e_3 + a_1 b_4 e_1 \varepsilon_4 + a_1 b_5 e_1 \varepsilon_5 + a_1 b_6 e_1 \varepsilon_6 + a_1 b_7 e_1 \varepsilon_7, \\
&+ a_2 b_0 e_2 + a_2 b_1 e_2 e_1 + a_2 b_2 e_2^2 + a_2 b_3 e_2 e_3 + a_2 b_4 e_2 \varepsilon_4 + a_2 b_5 e_2 \varepsilon_5 + a_2 b_6 e_2 \varepsilon_6 + a_2 b_7 e_2 \varepsilon_7, \\
&+ a_3 b_0 e_3 + a_3 b_1 e_3 e_1 + a_3 b_2 e_3 e_2 + a_3 b_3 e_3^2 + a_3 b_4 e_3 \varepsilon_4 + a_3 b_5 e_3 \varepsilon_5 + a_3 b_6 e_3 \varepsilon_6 + a_3 b_7 e_3 \varepsilon_7, \\
&+ a_4 b_0 \varepsilon_4 + a_4 b_1 \varepsilon_4 e_1 + a_4 b_2 \varepsilon_4 e_2 + a_4 b_3 \varepsilon_4 e_3 + a_4 b_4 \varepsilon_4^2 + a_4 b_5 \varepsilon_4 \varepsilon_5 + a_4 b_6 \varepsilon_4 \varepsilon_6 + a_4 b_7 \varepsilon_4 \varepsilon_7, \\
&+ a_5 b_0 \varepsilon_5 + a_5 b_1 \varepsilon_5 e_1 + a_5 b_2 \varepsilon_5 e_2 + a_5 b_3 \varepsilon_5 e_3 + a_5 b_4 \varepsilon_5 \varepsilon_4 + a_5 b_5 \varepsilon_5^2 + a_5 b_6 \varepsilon_5 \varepsilon_6 + a_5 b_7 \varepsilon_5 \varepsilon_7, \\
&+ a_6 b_0 \varepsilon_6 + a_6 b_1 \varepsilon_6 e_1 + a_6 b_2 \varepsilon_6 e_2 + a_6 b_3 \varepsilon_6 e_3 + a_6 b_4 \varepsilon_6 \varepsilon_4 + a_6 b_5 \varepsilon_6 \varepsilon_5 + a_6 b_6 \varepsilon_6^2 + a_6 b_7 \varepsilon_6 \varepsilon_7, \\
&+ a_7 b_0 \varepsilon_7 + a_7 b_1 \varepsilon_7 e_1 + a_7 b_2 \varepsilon_7 e_2 + a_7 b_3 \varepsilon_7 e_3 + a_7 b_4 \varepsilon_7 \varepsilon_4 + a_7 b_5 \varepsilon_7 \varepsilon_5 + a_7 b_6 \varepsilon_7 \varepsilon_6 + a_7 b_7 \varepsilon_7^2.
\end{aligned}$$

Then we have:

$$\begin{aligned}
y_0 &= x_0 b_0 - x_1 b_1 - x_2 b_2 - x_3 b_3 + x_4 b_4 + x_5 b_5 + x_6 p_6 + x_7 b_7, \\
y_1 &= x_0 b_1 + x_1 b_0 + x_2 b_3 - x_3 b_2 + x_4 b_5 - x_5 b_4 + x_6 b_7 - x_7 b_6, \\
y_2 &= x_0 b_2 - x_1 b_3 + x_2 b_0 + x_3 b_1 + x_4 b_6 - x_5 b_7 - x_6 b_4 + x_7 b_5, \\
y_3 &= x_0 b_3 + x_1 b_2 - x_2 b_1 + x_3 b_0 + x_4 b_7 + x_5 b_6 - x_6 b_5 - x_7 b_4, \\
y_4 &= x_0 b_4 + x_1 b_5 + x_2 b_6 + x_3 b_7 + x_4 b_0 - x_5 b_1 - x_6 b_2 - x_7 b_3, \\
y_5 &= x_0 b_5 + x_1 b_4 - x_2 b_7 + x_3 b_6 - x_4 b_1 + x_5 b_0 - x_6 b_3 + x_7 b_2, \\
y_6 &= x_0 b_6 + x_1 b_7 + x_2 b_4 - x_3 b_5 - x_4 b_2 + x_5 b_3 + x_6 b_0 - x_7 b_1, \\
y_7 &= x_0 b_7 - x_1 b_6 + x_2 b_5 + x_3 b_4 - x_4 b_3 - x_5 b_2 + x_6 b_1 + x_7 b_0.
\end{aligned}$$

We can see that the schoolbook method of multiplication of two hyperbolic octonions requires 64 real multiplications and 56 real additions.

In matrix notation, the above relations can be written more compactly as:

$$\mathbf{Y}_{8\times 1} = \mathbf{B}_8 \mathbf{X}_{8\times 1} \qquad (2)$$

where

$$\mathbf{X}_{8\times 1} = [x_0, x_1, x_2, x_3, x_4, x_5, x_6, x_7]^T, \quad \mathbf{Y}_{8\times 1} = [y_0, y_1, y_2, y_3, y_4, y_5, y_6, y_7]^T,$$

and

$$\mathbf{B}_8 = \left[\begin{array}{cccc|cccc} b_0 & -b_1 & -b_2 & -b_3 & b_4 & b_5 & b_6 & b_7 \\ b_1 & b_0 & b_3 & -b_2 & b_5 & -b_4 & b_7 & -b_6 \\ b_2 & -b_3 & b_0 & b_1 & b_6 & -b_7 & -b_4 & b_5 \\ b_3 & b_2 & -b_1 & b_0 & b_7 & b_6 & -b_5 & -b_4 \\ \hline b_4 & b_5 & b_6 & b_7 & b_0 & -b_1 & -b_2 & -b_3 \\ b_5 & b_4 & -b_7 & b_6 & -b_1 & b_0 & -b_3 & b_2 \\ b_6 & b_7 & b_4 & -b_5 & -b_2 & b_3 & b_0 & -b_1 \\ b_7 & -b_6 & b_5 & b_4 & -b_3 & -b_2 & b_1 & b_0 \end{array}\right],$$

The direct realization of (2) requires 64 real multiplications and 56 real additions too. We shall present the algorithm, which reduce arithmetical complexity to 26 real multiplications and 92 real additions.

## 3. Synthesis of a rationalized algorithm for multiplying two hyperbolic octonions

At first, we multiply by (-1) the sixth, seventh and eighth rows of the matrix $\mathbf{B}_8$. Then we interchange the first and the fifth column of this matrix and call the resulting matrix $\mathbf{B}'_8$.

$$\mathbf{B}'_8 = \left[\begin{array}{cccc|cccc} b_4 & -b_1 & -b_2 & -b_3 & b_0 & b_5 & b_6 & b_7 \\ b_5 & b_0 & b_3 & -b_2 & b_1 & -b_4 & b_7 & -b_6 \\ b_6 & -b_3 & b_0 & b_1 & b_2 & -b_7 & -b_4 & b_5 \\ b_7 & b_2 & -b_1 & b_0 & b_3 & b_6 & -b_5 & -b_4 \\ \hline b_0 & b_5 & b_6 & b_7 & b_4 & -b_1 & -b_2 & -b_3 \\ b_1 & -b_4 & b_7 & -b_6 & -b_5 & -b_0 & b_3 & -b_2 \\ b_2 & -b_7 & -b_4 & b_5 & -b_6 & -b_3 & -b_0 & b_1 \\ b_3 & b_6 & -b_5 & -b_4 & -b_7 & b_2 & -b_1 & -b_0 \end{array}\right].$$

Then we can write
$$\mathbf{Y}_{8\times 1} = \mathbf{R}_8^{(1)}\mathbf{B}'_8\mathbf{P}_8^{(1)}\mathbf{X}_{8\times 1}$$
where

$$\mathbf{P}_8^{(1)} = \left[\begin{array}{cccc|cccc} & & & & 1 & & & \\ & 1 & & & & & & \\ & & 1 & & & & & \\ & & & 1 & & & & \\ \hline 1 & & & & & & & \\ & & & & & 1 & & \\ & & & & & & 1 & \\ & & & & & & & 1 \end{array}\right], \quad \mathbf{R}_8^{(1)} = \left[\begin{array}{cccc|cccc} 1 & & & & & & & \\ & 1 & & & & & & \\ & & 1 & & & & & \\ & & & 1 & & & & \\ \hline & & & & 1 & & & \\ & & & & & -1 & & \\ & & & & & & -1 & \\ & & & & & & & -1 \end{array}\right].$$

This transformation is done in order to present a modified in this manner matrix as an algebraic sum of the block-symmetric Toeplitz-type matrix and some sparse matrix, i.e. matrix containing only small number of nonzero elements. Now the matrix $\mathbf{B}'_8$ can be represented as an algebraic sum of a symmetric Toeplitz-type matrix and another matrix which has many zero elements $\mathbf{B}'_8 = \mathbf{B}''_8 + 2\mathbf{M}_8^{(0)}$:

$$\mathbf{B}''_8 = \begin{bmatrix} b_4 & -b_1 & -b_2 & -b_3 & b_0 & b_5 & b_6 & b_7 \\ b_5 & -b_0 & b_3 & -b_2 & b_1 & -b_4 & b_7 & -b_6 \\ b_6 & -b_3 & -b_0 & b_1 & b_2 & -b_7 & -b_4 & b_5 \\ b_7 & b_2 & -b_1 & -b_0 & b_3 & b_6 & -b_5 & -b_4 \\ b_0 & b_5 & b_6 & b_7 & b_4 & -b_1 & -b_2 & -b_3 \\ b_1 & -b_4 & b_7 & -b_6 & b_5 & -b_0 & b_3 & -b_2 \\ b_2 & -b_7 & -b_4 & b_5 & b_6 & -b_3 & -b_0 & b_1 \\ b_3 & b_6 & -b_5 & -b_4 & b_7 & b_2 & -b_1 & -b_0 \end{bmatrix}, \quad \mathbf{M}_8^{(0)} = \begin{bmatrix} b_0 & & & & & & & \\ & b_0 & & & & \mathbf{0}_4 & & \\ & & b_0 & & & & & \\ & & & b_0 & & & & \\ & & & & & -b_5 & & \\ & \mathbf{0}_4 & & & & -b_6 & & \\ & & & & & -b_7 & & \end{bmatrix},$$

Taking into account a proposed decomposition, the computational procedure for multiplication hyperbolic octonions can be rewritten as follows:

$$\mathbf{Y}_{8\times 1} = \mathbf{R}_8 \mathbf{\Sigma}_{8\times 16} (\mathbf{B}''_8 \oplus 2\mathbf{M}_8^{(0)}) \mathbf{P}_{16\times 8}^{(2)} \mathbf{P}_8^{(1)} \mathbf{X}_{8\times 1} \qquad (3)$$

where sign "$\oplus$" – denotes the direct sum of two matrices [24],

$$\mathbf{R}_8 = diag(1,1,1,1,1,-1,-1,-1)$$

$$\mathbf{\Sigma}_{8\times 16} = \begin{bmatrix} 1 & & & & & 1 & & & & & & & & & & \\ & 1 & & & & & 1 & & & & & & & & & \\ & & 1 & & & & & 1 & & & & & & & & \\ & & & 1 & & & & & 1 & & & & & & & \\ & & & & 1 & & & & & 1 & & & & & & \\ & & & & & 1 & & & & & 1 & & & & & \\ & & & & & & 1 & & & & & 1 & & & & \\ & & & & & & & 1 & & & & & 1 & & & \end{bmatrix}, \quad \mathbf{P}_{16\times 8}^{(2)} = \begin{bmatrix} 1 & & & & & & & \\ & 1 & & & & & & \\ & & 1 & & & & & \\ & & & 1 & & & & \\ & & & & 1 & & & \\ & & & & & 1 & & \\ & & & & & & 1 & \\ & & & & & & & 1 \\ 1 & & & & & & & \\ & 1 & & & & & & \\ & & 1 & & & & & \\ & & & 1 & & & & \\ & & & & 1 & & & \\ & & & & & 1 & & \\ & & & & & & 1 & \\ & & & & & & & 1 \end{bmatrix}.$$

It is easy to see that $\mathbf{B}''_8$ has the following structure: $\mathbf{B}''_8 = \begin{bmatrix} \mathbf{A}_4 & \mathbf{B}_4 \\ \mathbf{B}_4 & \mathbf{A}_4 \end{bmatrix}$,

$$\mathbf{A}_4 = \begin{bmatrix} b_4 & -b_1 & -b_2 & -b_3 \\ b_5 & -b_0 & b_3 & -b_2 \\ b_6 & -b_3 & -b_0 & b_1 \\ b_7 & b_2 & -b_1 & -b_0 \end{bmatrix}, \quad \mathbf{B}_4 = \begin{bmatrix} b_0 & b_5 & b_6 & b_7 \\ b_1 & -b_4 & b_7 & -b_6 \\ b_2 & -b_7 & -b_4 & b_5 \\ b_3 & b_6 & -b_5 & -b_4 \end{bmatrix}.$$

It is easily to verify [25-27] that the matrix with this structure can be effectively factorized:

$$\mathbf{B}''_8 = (\mathbf{H}_2 \otimes \mathbf{I}_4) \frac{1}{2} [(\mathbf{A}_4 + \mathbf{B}_4) \oplus (\mathbf{A}_4 - \mathbf{B}_4)](\mathbf{H}_2 \otimes \mathbf{I}_4),$$

$$\mathbf{A}_4 + \mathbf{B}_4 = \left[\begin{array}{cc|cc} b_0+b_4 & -b_1+b_5 & -b_2+b_6 & -b_3+b_7 \\ b_1+b_5 & -b_0-b_4 & b_3+b_7 & -b_2-b_6 \\ \hline b_2+b_6 & -b_3-b_7 & -b_0-b_4 & b_1+b_5 \\ b_3+b_7 & b_2+b_6 & -b_1-b_5 & -b_0-b_4 \end{array}\right] = \mathbf{E}_4^{(0)},$$

$$\mathbf{A}_4 - \mathbf{B}_4 = \left[\begin{array}{cc|cc} -b_0+b_4 & -b_1-b_5 & -b_2-b_6 & -b_3-b_7 \\ -b_1+b_5 & -b_0+b_4 & b_3-b_7 & -b_2+b_6 \\ \hline -b_2+b_6 & -b_3+b_7 & -b_0+b_4 & b_1-b_5 \\ -b_3+b_7 & b_2-b_6 & -b_1+b_5 & -b_0+b_4 \end{array}\right] = \mathbf{F}_4^{(0)},$$

where $\mathbf{H}_2 = \begin{bmatrix} 1 & 1 \\ 1 & -1 \end{bmatrix}$ – is the order 2 Hadamard matrix, $\mathbf{I}_N$ is the order $N$ identity matrix, and sign "$\otimes$" denotes the Kronecker product of two matrices respectively [24].

Than the computational procedure for multiplication of the hyperbolic octonions at this step of the algorithm design can be represented as follows:

$$\mathbf{Y}_{8\times1} = \mathbf{R}_8 \mathbf{\Sigma}_{8\times16} \mathbf{W}_{16} \mathbf{D}_{16}^{(0)} \mathbf{W}_{16} \mathbf{P}_{16\times8}^{(2)} \mathbf{P}_8^{(1)} \mathbf{X}_{8\times1} \qquad (4)$$

where

$$\mathbf{W}_{16} = (\mathbf{H}_2 \otimes \mathbf{I}_4) \oplus \mathbf{I}_8 =$$

$$= \left[\begin{array}{cccc|cccc|cccccccc} 1 & & & & 1 & & & & & & & & & & & \\ & 1 & & & & 1 & & & & & & & & & & \\ & & 1 & & & & 1 & & & & & & & & & \\ & & & 1 & & & & 1 & & & & \mathbf{0}_8 & & & & \\ \hline 1 & & & & -1 & & & & & & & & & & & \\ & 1 & & & & -1 & & & & & & & & & & \\ & & 1 & & & & -1 & & & & & & & & & \\ & & & 1 & & & & -1 & & & & & & & & \\ \hline & & & & & & & & 1 & & & & & & & \\ & & & & & & & & & 1 & & & & & & \\ & & & & & & & & & & 1 & & & & & \\ & & & \mathbf{0}_8 & & & & & & & & 1 & & & & \\ & & & & & & & & & & & & 1 & & & \\ & & & & & & & & & & & & & 1 & & \\ & & & & & & & & & & & & & & 1 & \\ & & & & & & & & & & & & & & & 1 \end{array}\right],$$

$$\mathbf{D}_{16}^{(0)} = (\mathbf{E}_4^{(0)} \oplus \mathbf{F}_4^{(0)}) \oplus 2\mathbf{M}_8^{(0)} = \left[\begin{array}{cc|c} \frac{1}{2}\mathbf{E}_4^{(0)} & \mathbf{0}_4 & \\ \mathbf{0}_4 & \frac{1}{2}\mathbf{F}_4^{(1)} & \mathbf{0}_8 \\ \hline & \mathbf{0}_8 & 2\mathbf{M}_8^{(0)} \end{array}\right].$$

Fig. 1 shows a data flow diagram of the rationalized algorithm for computation of a product of a hyperbolic octonions. In this paper, data flow diagrams are oriented from left to right. Straight lines in the figures denote the operations of data transfer. Points where lines converge denote summation. The dashed lines indicate the sign change operation. We deliberately use the usual lines without arrows on purpose, so as not to clutter the picture. The rectangles indicate the matrix–vector multiplications with the matrix inscribed inside a rectangle.

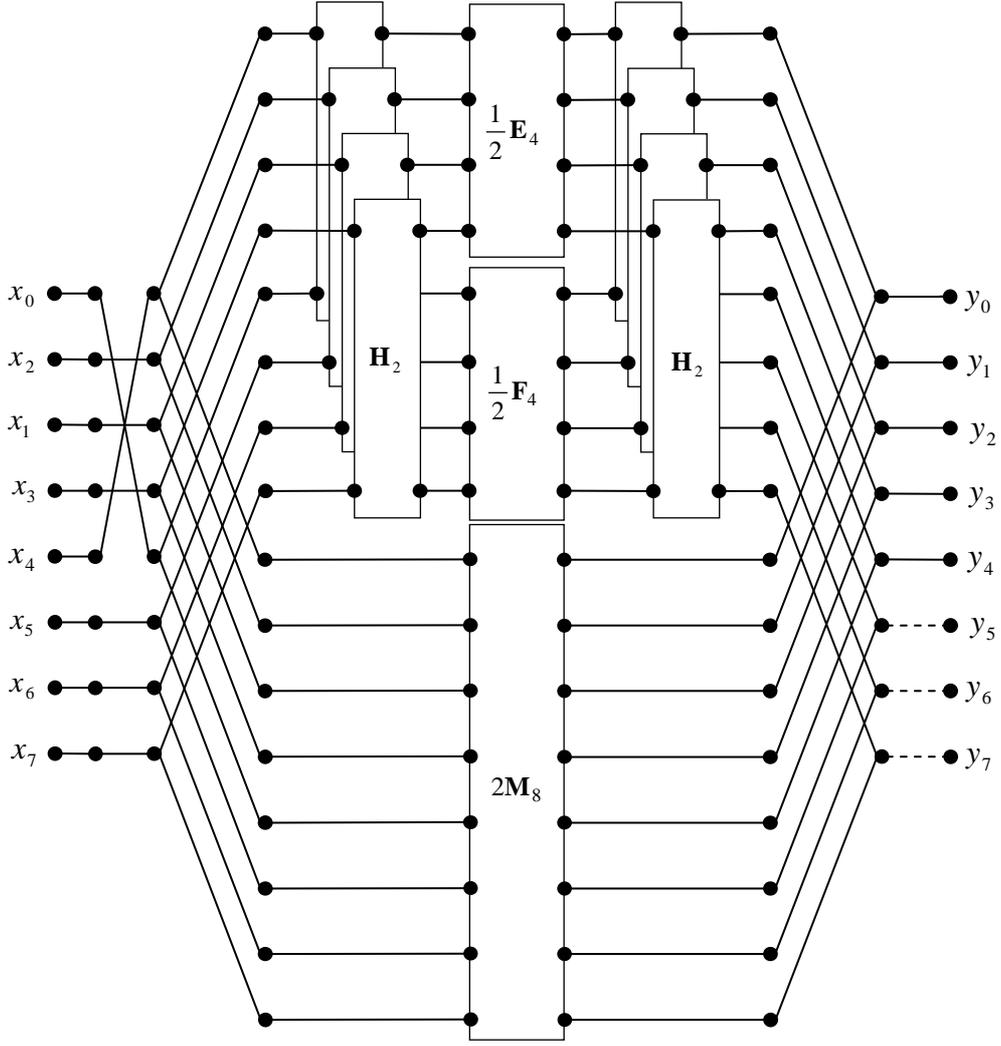

**Fig. 1.** Data flow diagram for rationalized hyperbolic octonion multiplication algorithm in accordance with the procedure (4).

Let us now consider the structures of the matrices $\mathbf{E}_4^{(0)}$ and $\mathbf{F}_4^{(0)}$. First we multiply by (-1) every element of the first row of matrix $\mathbf{E}_4^{(0)}$ and call the resulting matrix $\mathbf{E}_4^{(1)}$:

$$\mathbf{E}_4^{(1)} = \left[\begin{array}{cc|cc} -b_0-b_4 & b_1-b_5 & b_2-b_6 & b_3-b_7 \\ b_1+b_5 & -b_0-b_4 & b_3+b_7 & -b_2-b_6 \\ \hline b_2+b_6 & -b_3-b_7 & -b_0-b_4 & b_1+b_5 \\ b_3+b_7 & b_2+b_6 & -b_1-b_5 & -b_0-b_4 \end{array}\right].$$

The matrix $\mathbf{E}_4^{(1)}$ can be decomposed as an algebraic sum of a symmetric Toeplitz matrix and another matrix which has many zero elements $\mathbf{E}_4^{(1)} = \mathbf{E}_4^{(2)} + 2\mathbf{M}_4^{(1)}$:

$$\mathbf{E}_4^{(1)} = \left[\begin{array}{cc|cc} -b_0-b_4 & b_1+b_5 & b_2+b_6 & b_3+b_7 \\ b_1+b_5 & -b_0-b_4 & b_3+b_7 & b_2+b_6 \\ \hline b_2+b_6 & b_3+b_7 & -b_0-b_4 & b_1+b_5 \\ b_3+b_7 & b_2+b_6 & b_1+b_5 & -b_0-b_4 \end{array}\right] + 2\left[\begin{array}{cc|cc} 0 & -b_5 & -b_6 & -b_7 \\ 0 & 0 & 0 & -b_2-b_6 \\ \hline 0 & -b_3-b_7 & 0 & 0 \\ 0 & 0 & -b_1-b_5 & 0 \end{array}\right],$$

$$\mathbf{E}_4^{(2)} = \begin{bmatrix} -b_0-b_4 & b_1+b_5 & b_2+b_6 & b_3+b_7 \\ b_1+b_5 & -b_0-b_4 & b_3+b_7 & b_2+b_6 \\ \hline b_2+b_6 & b_3+b_7 & -b_0-b_4 & b_1+b_5 \\ b_3+b_7 & b_2+b_6 & b_1+b_5 & -b_0-b_4 \end{bmatrix} = \begin{bmatrix} \mathbf{A}_2 & \mathbf{B}_2 \\ \hline \mathbf{B}_2 & \mathbf{A}_2 \end{bmatrix},$$

$$\mathbf{A}_2 = \begin{bmatrix} -b_0-b_4 & b_1+b_5 \\ \hline b_1+b_5 & -b_0-b_4 \end{bmatrix}, \quad \mathbf{B}_2 = \begin{bmatrix} b_2+b_6 & b_3+b_7 \\ \hline b_3+b_7 & b_2+b_6 \end{bmatrix},$$

$$\mathbf{M}_4^{(1)} = \begin{bmatrix} 0 & -b_5 & -b_6 & -b_7 \\ 0 & 0 & 0 & -b_2-b_6 \\ \hline 0 & -b_3-b_7 & 0 & 0 \\ 0 & 0 & -b_1-b_5 & 0 \end{bmatrix} \quad (5)$$

$$\mathbf{E}_4^{(2)} = (\mathbf{H}_2 \otimes \mathbf{I}_2) \frac{1}{2}[(\mathbf{A}_2 + \mathbf{B}_2) \oplus (\mathbf{A}_2 - \mathbf{B}_2)](\mathbf{H}_2 \otimes \mathbf{I}_2) \quad (6)$$

$$\mathbf{A}_2 + \mathbf{B}_2 = \begin{bmatrix} -b_0-b_4+b_2+b_6 & b_1+b_5+b_3+b_7 \\ \hline b_1+b_5+b_3+b_7 & -b_0-b_4+b_2+b_6 \end{bmatrix} = \mathbf{E}_2^{(0)},$$

$$\mathbf{A}_2 - \mathbf{B}_2 = \begin{bmatrix} -b_0-b_4-b_2-b_6 & b_1+b_5-b_3-b_7 \\ \hline b_1+b_5-b_3-b_7 & -b_0-b_4-b_2-b_6 \end{bmatrix} = \mathbf{F}_2^{(0)}.$$

Let us return now to the structure of the matrix $\mathbf{F}_4^{(0)}$. It easy to see than the matrix $\mathbf{F}_4^{(0)}$ can also be represented as an algebraic sum of a symmetric Toeplitz matrix and another matrix which has many zero elements $\mathbf{F}_4^{(0)} = \mathbf{F}_4^{(1)} + 2\mathbf{M}_4^{(2)}$:

$$\mathbf{F}_4^{(1)} = \begin{bmatrix} -b_0+b_4 & -b_1+b_5 & -b_2+b_6 & -b_3+b_7 \\ -b_1+b_5 & -b_0+b_4 & -b_3+b_7 & -b_2+b_6 \\ \hline -b_2+b_6 & -b_3+b_7 & -b_0+b_4 & -b_1+b_5 \\ -b_3+b_7 & -b_2+b_6 & -b_1+b_5 & -b_0+b_4 \end{bmatrix} + 2\begin{bmatrix} 0 & -b_5 & -b_5 & -b_5 \\ 0 & 0 & b_3-b_7 & 0 \\ \hline 0 & 0 & 0 & b_1-b_5 \\ 0 & b_2-b_6 & 0 & 0 \end{bmatrix},$$

$$\mathbf{F}_4^{(1)} = \begin{bmatrix} -b_0+b_4 & -b_1+b_5 & -b_2+b_6 & -b_3+b_7 \\ -b_1+b_5 & -b_0+b_4 & -b_3+b_7 & -b_2+b_6 \\ \hline -b_2+b_6 & -b_3+b_7 & -b_0+b_4 & -b_1+b_5 \\ -b_3+b_7 & -b_2+b_6 & -b_1+b_5 & -b_0+b_4 \end{bmatrix} = \begin{bmatrix} \mathbf{C}_2 & \mathbf{D}_2 \\ \hline \mathbf{D}_2 & \mathbf{C}_2 \end{bmatrix},$$

$$\mathbf{C}_2 = \begin{bmatrix} -b_0+b_4 & -b_1+b_5 \\ \hline -b_1+b_5 & -b_0+b_4 \end{bmatrix}, \quad \mathbf{D}_2 = \begin{bmatrix} -b_2+b_6 & -b_3+b_7 \\ \hline -b_3+b_7 & -b_2+b_6 \end{bmatrix},$$

$$\mathbf{M}_4^{(2)} = \begin{bmatrix} 0 & -b_5 & -b_5 & -b_5 \\ 0 & 0 & b_3-b_7 & 0 \\ \hline 0 & 0 & 0 & b_1-b_5 \\ 0 & b_2-b_6 & 0 & 0 \end{bmatrix} \quad (7)$$

It is easily to verify [25-27] that the matrix $\mathbf{F}_4^{(1)}$ can be factorized in the same way:

$$\mathbf{F}_4^{(1)} = (\mathbf{H}_2 \otimes \mathbf{I}_2) \frac{1}{2}[(\mathbf{C}_2 + \mathbf{D}_2) \oplus (\mathbf{C}_2 - \mathbf{D}_2)](\mathbf{H}_2 \otimes \mathbf{I}_2) \quad (8)$$

$$\mathbf{C}_2 + \mathbf{D}_2 = \left[\begin{array}{c|c} -b_0+b_4-b_2+b_6 & -b_1+b_5-b_3+b_7 \\ \hline -b_1+b_5-b_3+b_7 & -b_0+b_4-b_2+b_6 \end{array}\right] = \mathbf{K}_2^{(0)},$$

$$\mathbf{C}_2 - \mathbf{D}_2 = \left[\begin{array}{c|c} -b_0+b_4+b_2-b_6 & -b_1+b_5+b_3-b_7 \\ \hline -b_1+b_5+b_3-b_7 & -b_0+b_4+b_2-b_6 \end{array}\right] = \mathbf{L}_2^{(0)}.$$

Substituting (5), (6), (7), and (8) in (4) we can write:

$$\mathbf{Y}_{8\times 1} = \mathbf{R}_8 \mathbf{\Sigma}_{8\times 16}^{(3)} \mathbf{W}_{16} \mathbf{R}_{16} \mathbf{\Sigma}_{16\times 24}^{(5)} \mathbf{W}_{24}^{(1)} \mathbf{D}_{24}^{(1)} \mathbf{W}_{24}^{(1)} \mathbf{P}_{24\times 16}^{(4)} \mathbf{W}_{16} \mathbf{P}_{16\times 8}^{(2)} \mathbf{P}_8^{(1)} \mathbf{X}_{8\times 1} \qquad (9)$$

where

$$\mathbf{\Sigma}_{16\times 24}^{(5)} = \begin{bmatrix}
1 & & & & 1 & & & & & & & & & & & & & & & & & & & \\
 & 1 & & & & 1 & & & & & & & & & & & & & & & & & & \\
 & & 1 & & & & 1 & & & & & & & & & & & & & & & & & \\
 & & & 1 & & & & 1 & & & & & & & & & & & & & & & & \\
 & & & & & & & & 1 & & & & 1 & & & & & & & & & & & \\
 & & & & & & & & & 1 & & & & 1 & & & & & & & & & & \\
 & & & & & & & & & & 1 & & & & 1 & & & & & & & & & \\
 & & & & & & & & & & & 1 & & & & 1 & & & & & & & & \\
 & & & & & & & & & & & & & & & & 1 & & & & & & & \\
 & & & & & & & & & & & & & & & & & 1 & & & & & & \\
 & & & & & & & & & & & & & & & & & & 1 & & & & & \\
 & & & & & & & & & & & & & & & & & & & 1 & & & & \\
 & & & & & & & & & & & & & & & & & & & & 1 & & & \\
 & & & & & & & & & & & & & & & & & & & & & 1 & & \\
 & & & & & & & & & & & & & & & & & & & & & & 1 & \\
 & & & & & & & & & & & & & & & & & & & & & & & 1
\end{bmatrix}.$$

$$\mathbf{D}_{24}^{(1)} = \mathbf{Q}_8^{(2)} \oplus \mathbf{Q}_8^{(3)} \oplus 2\mathbf{M}_8, \quad \mathbf{Q}_8^{(2)} = \frac{1}{4}(\mathbf{E}_2^{(0)} \oplus \mathbf{F}_2^{(0)}) \oplus \mathbf{M}_4^{(1)}, \quad \mathbf{Q}_8^{(3)} = \frac{1}{4}(\mathbf{K}_2^{(0)} \oplus \mathbf{L}_2^{(0)}) \oplus \mathbf{M}_4^{(2)},$$

$$\mathbf{Q}_8^{(2)} = \left[\begin{array}{cc|c} \frac{1}{4}\mathbf{E}_2^{(0)} & \mathbf{0}_2 & \mathbf{0}_4 \\ \mathbf{0}_2 & \frac{1}{4}\mathbf{F}_2^{(0)} & \\ \hline \mathbf{0}_4 & & \mathbf{M}_4^{(1)} \end{array}\right],$$

$$\mathbf{Q}_8^{(3)} = \left[\begin{array}{cc|c} \frac{1}{4}\mathbf{K}_2^{(0)} & \mathbf{0}_2 & \mathbf{0}_4 \\ \mathbf{0}_2 & \frac{1}{4}\mathbf{L}_2^{(0)} & \\ \hline \mathbf{0}_4 & & \mathbf{M}_4^{(2)} \end{array}\right].$$

$$\mathbf{W}_{24}^{(1)} = (\mathbf{H}_2 \otimes \mathbf{I}_2) \oplus \mathbf{I}_4 \oplus (\mathbf{H}_2 \otimes \mathbf{I}_2) \oplus \mathbf{I}_{12} =$$

$$\mathbf{W}_{24}^{(1)} = \begin{bmatrix} \begin{array}{cccc|cccc} 1 & 1 & & & & & & \\ & & 1 & 1 & & \mathbf{0}_4 & & \\ 1 & -1 & & & & & & \\ & & 1 & -1 & & & & \\ \hline & & & & 1 & & & \\ & \mathbf{0}_4 & & & & 1 & & \\ & & & & & & 1 & \\ & & & & & & & 1 \end{array} & \mathbf{0}_8 & \mathbf{0}_8 \\ \mathbf{0}_8 & \begin{array}{cccc|cccc} 1 & 1 & & & & & & \\ & & 1 & 1 & & \mathbf{0}_4 & & \\ 1 & -1 & & & & & & \\ & & 1 & -1 & & & & \\ \hline & & & & 1 & & & \\ & \mathbf{0}_4 & & & & 1 & & \\ & & & & & & 1 & \\ & & & & & & & 1 \end{array} & \mathbf{0}_8 \\ \mathbf{0}_8 & \mathbf{0}_8 & \begin{array}{cccccccc} 1 & & & & & & & \\ & 1 & & & & & & \\ & & 1 & & & & & \\ & & & 1 & & & & \\ & & & & 1 & & & \\ & & & & & 1 & & \\ & & & & & & 1 & \\ & & & & & & & 1 \end{array} \end{bmatrix},$$

$$\mathbf{P}_{24 \times 16}^{(4)} = \begin{bmatrix} \begin{array}{cccc} 1 & & & \\ & 1 & & \\ & & 1 & \\ & & & 1 \end{array} & & & \\ \hline \begin{array}{cccc} 1 & & & \\ & 1 & & \\ & & 1 & \\ & & & 1 \end{array} & & & \\ \hline & \begin{array}{cccc} 1 & & & \\ & 1 & & \\ & & 1 & \\ & & & 1 \end{array} & & \\ \hline & \begin{array}{cccc} 1 & & & \\ & 1 & & \\ & & 1 & \\ & & & 1 \end{array} & & \\ \hline & & \begin{array}{cccccccc} 1 & & & & & & & \\ & 1 & & & & & & \\ & & 1 & & & & & \\ & & & 1 & & & & \\ & & & & 1 & & & \\ & & & & & 1 & & \\ & & & & & & 1 & \\ & & & & & & & 1 \end{array} \end{bmatrix}$$

Fig. 2 shows a data flow diagram of the rationalized algorithm for multiplying of two hyperbolic octonions at the second stage of synthesis.

Consider now the matrices $\mathbf{E}_2^{(0)}$, $\mathbf{F}_2^{(0)}$, $\mathbf{K}_2^{(0)}$, and $\mathbf{L}_2^{(0)}$. As can be seen, these matrices also have a "good" structures leading to a decrease in the number of real multiplications during calculation of the hyperbolic octonion product.

$$\mathbf{E}_2^{(0)} = \left[\begin{array}{c|c} -b_0-b_4+b_2+b_6 & b_1+b_5+b_3+b_7 \\ \hline b_1+b_5+b_3+b_7 & -b_0-b_4+b_2+b_6 \end{array}\right] = \left[\begin{array}{c|c} a & b \\ \hline b & a \end{array}\right] = \mathbf{H}_2 \frac{1}{2}[(a+b) \oplus (a-b)]\mathbf{H}_2,$$

$$\mathbf{F}_2^{(0)} = \left[\begin{array}{c|c} -b_0-b_4-b_2-b_6 & b_1+b_5-b_3-b_7 \\ \hline b_1+b_5-b_3-b_7 & -b_0-b_4-b_2-b_6 \end{array}\right] = \left[\begin{array}{c|c} c & d \\ \hline d & c \end{array}\right] = \mathbf{H}_2 \frac{1}{2}[(c+d) \oplus (c-d)]\mathbf{H}_2,$$

$$\mathbf{K}_2^{(0)} = \left[\begin{array}{c|c} -b_0+b_4-b_2+b_6 & -b_1+b_5-b_3+b_7 \\ \hline -b_1+b_5-b_3+b_7 & -b_0+b_4-b_2+b_6 \end{array}\right] = \left[\begin{array}{c|c} e & f \\ \hline f & e \end{array}\right] = \mathbf{H}_2 \frac{1}{2}[(e+f) \oplus (e-f)]\mathbf{H}_2,$$

$$\mathbf{L}_2^{(0)} = \left[\begin{array}{c|c} -b_0+b_4+b_2-b_6 & -b_1+b_5+b_3-b_7 \\ \hline -b_1+b_5+b_3-b_7 & -b_0+b_4+b_2-b_6 \end{array}\right] = \left[\begin{array}{c|c} g & h \\ \hline h & g \end{array}\right] = \mathbf{H}_2 \frac{1}{2}[(g+h) \oplus (g-h)]\mathbf{H}_2.$$

Introduce the following notation:

$$a+b = c_0 = -b_0-b_4+b_2+b_6+b_1+b_5+b_3+b_7, \quad a-b = c_1 = -b_0-b_4+b_2+b_6-b_1-b_5-b_3-b_7,$$

$$c+d = c_2 = -b_0-b_4-b_2-b_6+b_1+b_5-b_3-b_7, \quad c-d = c_3 = -b_0-b_4-b_2-b_6-b_1-b_5+b_3+b_7,$$

$$e+f = c_4 = -b_0+b_4-b_2+b_6-b_1+b_5-b_3+b_7, \quad e-f = c_5 = -b_0+b_4-b_2+b_6+b_1-b_5+b_3-b_7,$$

$$g+h = c_6 = -b_0+b_4+b_2-b_6-b_1+b_5+b_3-b_7, \quad g-h = c_7 = -b_0+b_4+b_2-b_6+b_1-b_5-b_3+b_7.$$

and

$$s_0 = \frac{1}{8}c_0, \ s_1 = \frac{1}{8}c_1, \ s_2 = \frac{1}{8}c_2, \ s_3 = \frac{1}{8}c_3, \ s_4 = \frac{1}{8}c_4, \ s_5 = \frac{1}{8}c_5, \ s_6 = \frac{1}{8}c_6, \ s_7 = \frac{1}{8}c_7.$$

Using the above notations and combining partial decompositions in a single computational procedure we finally can write following:

$$\mathbf{Y}_{8\times 1} = \mathbf{R}_8 \mathbf{P}_{8\times 16}^{(3)} \mathbf{W}_{16} \mathbf{R}_{16} \mathbf{\Sigma}_{16\times 24}^{(5)} \mathbf{W}_{24}^{(1)} \mathbf{W}_{24}^{(2)} \mathbf{D}_{24}^{(2)} \mathbf{W}_{24}^{(2)} \mathbf{W}_{24}^{(1)} \mathbf{P}_{24\times 16}^{(4)} \mathbf{W}_{16} \mathbf{P}_{16\times 8}^{(2)} \mathbf{P}_8^{(1)} \mathbf{X}_{8\times 1} \qquad (10)$$

where

$$\mathbf{D}_{24}^{(2)} = \mathbf{Q}_8^{(4)} \oplus \mathbf{Q}_8^{(5)} \oplus 2\mathbf{M}_8, \quad \mathbf{Q}_8^{(4)} = s_0 \oplus s_1 \oplus s_2 \oplus s_3 \oplus \mathbf{M}_4^{(1)}, \quad \mathbf{Q}_8^{(5)} = s_4 \oplus s_5 \oplus s_6 \oplus s_7 \oplus \mathbf{M}_4^{(2)}.$$

$$\mathbf{Q}_8^{(4)} = \left[\begin{array}{cc|cc} s_0 & 0 & & \\ 0 & s_1 & \mathbf{0}_2 & \mathbf{0}_4 \\ \hline \mathbf{0}_2 & & s_2 & 0 \\ & & 0 & s_3 \\ \hline \mathbf{0}_4 & & & \mathbf{M}_4^{(1)} \end{array}\right], \quad \mathbf{Q}_8^{(5)} = \left[\begin{array}{cc|cc} s_4 & 0 & & \\ 0 & s_5 & \mathbf{0}_2 & \mathbf{0}_4 \\ \hline \mathbf{0}_2 & & s_6 & 0 \\ & & 0 & s_7 \\ \hline \mathbf{0}_4 & & & \mathbf{M}_4^{(2)} \end{array}\right],$$

$$\mathbf{W}_{24}^{(2)} = (\mathbf{I}_2 \otimes \mathbf{H}_2) \oplus \mathbf{I}_4 \oplus (\mathbf{I}_2 \otimes \mathbf{H}_2) \oplus \mathbf{I}_{12} =$$

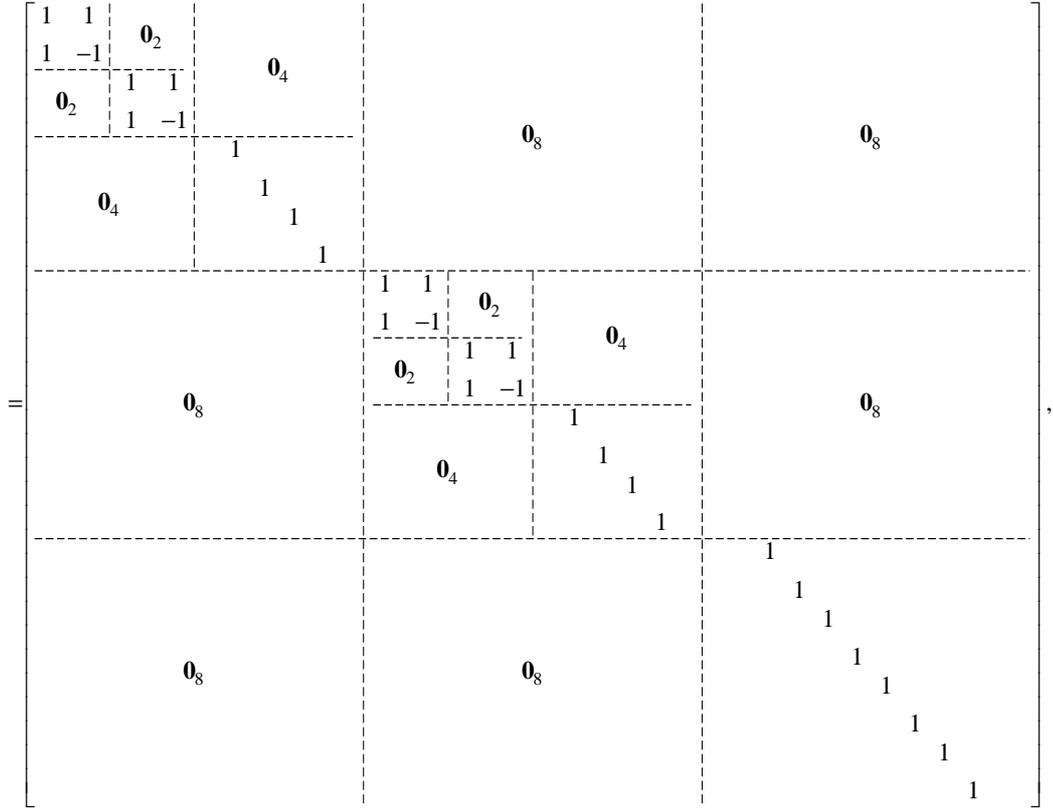

Fig. 3 shows a data flow diagram of the rationalized algorithm for multiplying of two hyperbolic octonions at the final stage of the algorithm derivation. The circles in this figure show the operation of multiplication by a variable (or constant) inscribed inside a circle.

We can see that the ordinary approach to calculation of elements $\{s_k\}, k=0,1,...,7$ requires 56 additions. It is easy to see that the relations for calculation of $\{s_k\}$ contain repeated algebraic sums. Therefore, the number of additions necessary to calculate these elements can be reduced. So, it is easy to verify that the elements $\{s_k\}$, $k=0,1,...,7$ can be calculated using the following rationalized matrix–vector procedure:

$$\mathbf{S}_{8\times 1} = \frac{1}{8}\mathbf{P}_8^{(7)}\mathbf{P}_8^{(6)}\mathbf{W}_8\mathbf{R}_8^{(1)}\mathbf{B}_{8\times 1} \qquad (11)$$

$$\mathbf{W}_8 = \mathbf{H}_2 \otimes \mathbf{I}_4,$$

$\mathbf{S}_{8\times 1} = [s_0, s_1, s_2, s_3, s_4, s_5, s_6, s_7]^T$, $\mathbf{B}_{8\times 1} = [b_0, b_1, b_2, b_3, b_4, b_5, b_6, b_7]^T$, $\mathbf{R}_8^{(1)} = diag(-1,1,1,1,1,1,1,1)$.

$$\mathbf{P}_8^{(6)} = \begin{bmatrix} 1 & & & & 1 & & & \\ & 1 & & 1 & & & & \\ & & 1 & & 1 & & & \\ & 1 & & -1 & & & & \\ & & -1 & & 1 & & & \\ & & & & & 1 & & 1 \\ 1 & & & & & & -1 & \\ & & & & & 1 & & -1 \end{bmatrix}, \quad \mathbf{P}_8^{(7)} = \begin{bmatrix} & 1 & 1 & & & & & \\ & -1 & 1 & & & & & \\ & & & & 1 & 1 & & \\ & & & & -1 & 1 & & \\ & & & -1 & 1 & & & \\ & & & 1 & 1 & & & \\ 1 & & & & & & & -1 \\ 1 & & & & & & & 1 \end{bmatrix}.$$

Fig. 4 shows a data flow diagram of the process for calculating the vector $\mathbf{S}_{8\times 1}$ elements.

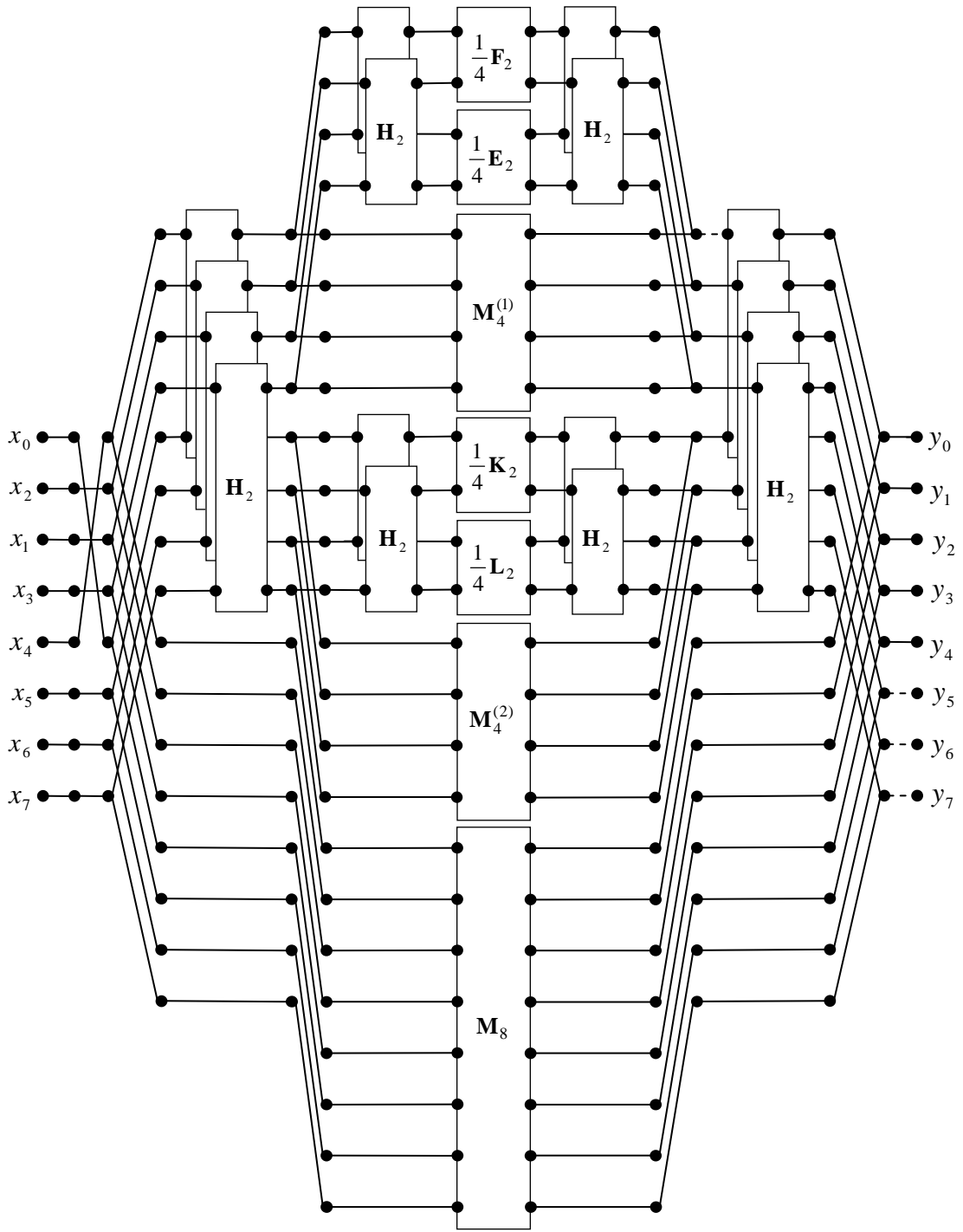

**Fig. 2.** Data flow diagram for rationalized hyperbolic octonion multiplication algorithm in accordance with the procedure (9).

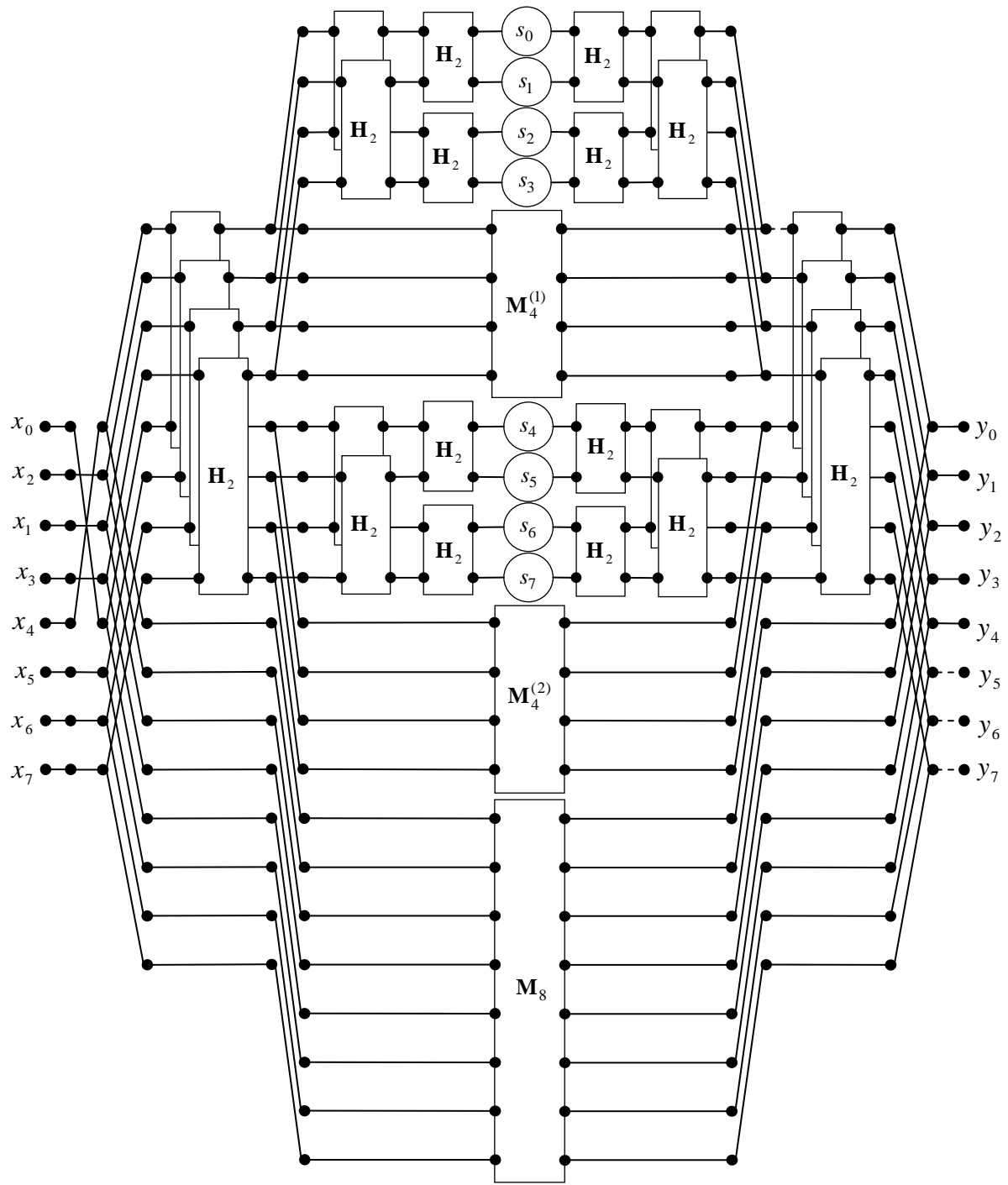

**Fig. 3.** Data flow diagram for rationalized hyperbolic octonion multiplication algorithm in accordance with the procedure (10).

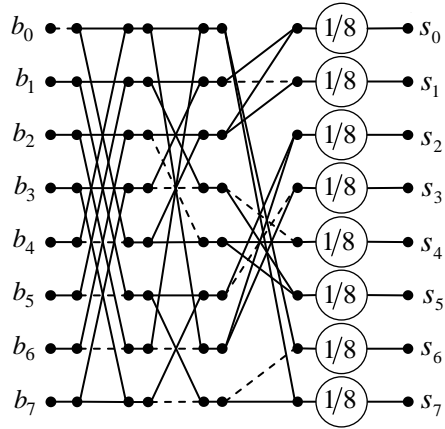

**Fig. 4.** Data flow diagram describing the process of calculating elements of the vector $\mathbf{S}_{8\times 1}$ in accordance with the procedure (11).

## 4. Estimation of computational complexity

We calculate how many real multiplications (excluding multiplications by power of two) and real additions are required for realization of the proposed algorithm, and compare it with the number of operations required for a direct evaluation of matrix-vector product in Eq. (2). Let us look to the data flow diagram in Figure 3. It is easy to verify that all the real multiplications which to be performed to computing the product of two hyperbolic octonions are realized only during multiplying a vector of data by the quasi-diagonal matrix $\mathbf{D}_{24}^{(2)}$. It can be argued that the multiplication of a vector by the matrix $\mathbf{D}_{24}^{(2)}$ requires 26 real multiplications and only a few trivial multiplications by the power of two. Multiplication by power of two may be implemented using convention arithmetic shift operations, which have simple realization and hence may be neglected during computational complexity estimation. So, the number of real multiplications required using the proposed algorithm is 26. Thus using the proposed algorithm the number of real multiplications to calculate the hyperbolic octonion product is significantly reduced.

Now we calculate the number of additions required in the implementation of the algorithm. It is easily to verify that the number of real additions required using our algorithm is 92. Therefore, the total number of arithmetic operations is still slightly less than the total number of arithmetic operations in the naive algorithm.

## 5. Conclusion

In this paper, we have presented an original algorithm that allows us to compute the product of two hyperbolic octonions with reduced multiplicative complexity. The proposed algorithm saves 38 real multiplications compared to the schoolbook algorithm. Unfortunately, the number of real additions in the proposed algorithm is somewhat greater than in the direct algorithm, but the total number of arithmetical operations is still less. For applications where the "cost" of a real multiplication is greater than that of a real addition, the new algorithm is generally more efficient than direct method.